\documentclass{article}

\usepackage{arxiv}

\usepackage[utf8]{inputenc} 
\usepackage[T1]{fontenc}    
\usepackage{hyperref}       
\usepackage{url}            
\usepackage{booktabs}       
\usepackage{amsfonts}       
\usepackage{nicefrac}       
\usepackage{microtype}      
\usepackage{lipsum}
\usepackage{amsmath}
\usepackage{threeparttable}

\title{On a scalable problem transformation method for multi-label learning}

\author{
        Dora Jambor \\
        Shopify Inc. \\
        \texttt{dora.jambor@shopify.com} \\
   \And
        Peng Yu \\
        Shopify Inc. \\
        \texttt{peng.yu@shopify.com} \\
}

\begin{document}
\maketitle

\begin{abstract}
Binary relevance is a simple approach to solve multi-label learning problems where an independent binary classifier is built per each label. A common challenge with this in real-world applications is that the label space can be very large, making it difficult to use binary relevance to larger scale problems. In this paper, we propose a scalable alternative to this, via transforming the multi-label problem into a single binary classification. We experiment with a few variations of our method and show that our method achieves higher precision than binary relevance and faster execution times on a top-K recommender system task.
\end{abstract}


\section{Introduction}
A simple and performant approach commonly used in real-world multi-label learning applications is Binary Relevance (BR) \cite{tsoumakas2007multi}. BR is a decomposition method that trains a single binary classifier for each label to classify input instances as relevant or irrelevant for the given label.

As the number of label sets grows exponentially with increases in the number of class labels, a key challenge in multi-label learning is scalability. In the BR framework, when training individual binary classifiers for each label sequentially prohibitively takes a long time, we might resort to parallelising our models across workers. However, this approach increases the I/O cost of reading the input data and transferring the data across the network. In this work, we propose a set of transformations which allows us to solve a multi-label learning problem by solving one binary classification problem. We demonstrate the effectiveness of our proposed method via a real-world top-K recommendation task and show that we are able to solve our problem much faster, and with higher precision.

\section{Background and Methodology}

As discussed by \cite{tsoumakas2009mining}, there are two main paradigms to solve multi-label problems: (i) \textit{problem transformation}, (ii) \textit{algorithm adaptation}. The former transforms the learning task into one or more single-label classification tasks, whereas the latter extends specific learning algorithms in order to handle multi-label data directly. Similar to BR, our proposed method is an instance of (i).

Formally, let $X \in\mathbb{R}^{m\times n}$ be a matrix of $m$ instances where each instance is an $n$-dimensional feature vector, and let $Y \in\{0,1\}^{m\times k}$ be a matrix of $m$ responses where each response if a $k$-dimensional label vector.

Inspired by Kesler’s construction  \cite{nilsson1965learning, duda1973pattern} , an approach to extend learning algorithms for binary classification to the multiclass case \cite{har2003constraint} , we propose a set of transformations on $X$ and $Y$ to convert the multi-label problem into a single binary classification problem. We also introduce a way to map the solutions of the binary classification problem back to the original multi-label setting. These transformations are defined as:

\begin{center}
 $$X'=diag(\underbrace{X, \ldots, X}_k)    Y'= \left[\begin{smallmatrix} 
                                                            Y_1\\ 
                                                            \cdots \\
                                                            Y_k
                                                    \end{smallmatrix} \right]
                                                    $$
\end{center} 
where $Y_1, \ldots, Y_k$ are $m\times 1$-dimensional vectors corresponding to $m$ responses for each label, i.e. $Y = \left[\begin{smallmatrix} 
                                                            Y_1 & \cdots & Y_k
                                                    \end{smallmatrix} \right]$.

Using the transformed $X'\in \mathbb{R}^{mk \times nk}$ and $Y'\in \{0,1\}^{mk \times 1}$, we solve a single binary classification with $X'$ as instances and $Y'$ as responses. After obtaining $\widehat{Y'} = \left[\begin{smallmatrix} 
                                                            \widehat{Y'}_1\\ 
                                                            \cdots \\
                                                            \widehat{Y'}_k
                                                    \end{smallmatrix} \right]$, our estimates of $Y'$, we assign $\widehat{Y} = \left[\begin{smallmatrix} 
\widehat{Y'}_1 & \cdots & \widehat{Y'}_k
\end{smallmatrix} \right]$ as the predicted label scores of the original multi-label problem.

\section{Experiments and discussion}
We conduct our experiments on an internal dataset containing 705,093 users’ app installations for the top 100 most popular apps \footnote{Items in a recommendation task constitute as the label set in a multi-label learning problem setting.  Top-K recommendations for every user is obtained by picking the top K label estimates per user.} on Shopify App Store \footnote{Shopify App Store: http://apps.shopify.com}. In this task, both $X$ and $Y$ are the binary user-item matrix composed of each user's historical app installations.  We then perform a three fold time-series based split to obtain three pairs of train-test dataset. \footnote{We performed the temporal split such that all user-item interactions in the training set were interactions that happened before interactions contained in the test set.}

We compare our method (termed DiagT) against BR as a baseline.  We seek to answer if our approach is amenable to the application of dimensionality reduction techniques due to the sparsity and large size of $X'$ by investigating the performance of a few variations of DiagT, utilizing the hashing trick  \cite{langford2007vowpal}  and random undersampling \footnote{https://imbalanced-learn.readthedocs.io}.
\begin{table}[b]
\label{table:results}
\caption{Summary of experiments}
\begin{tabular}{lllllllll}
\toprule
\textbf{models}     & \multicolumn{1}{l}{\textbf{\# nnz}} & \multicolumn{1}{l}{\textbf{density (\%)}} & \multicolumn{1}{l}{\textbf{speed (s)}} & \textbf{p@1 (\%)} & \multicolumn{1}{l}{\textbf{p@5 (\%)}} & \multicolumn{1}{l}{\textbf{p@10 (\%)}} \\
\midrule
    BR         & 2,594,150                  & 5                                & 216.4                         & 17.9 $\pm$ 1.0 & 18.2 $\pm$ 1.4  & 20.2 $\pm$ 1.2    \\
DiagT      & 273,942,306                & 0.05                             & \textbf{172.3}                         & \textbf{21.7} $\pm$ 1.0                   & \textbf{21.4} $\pm$ 1.6                     & \textbf{23.6} $\pm$ 1.2                      \\
DiagT-hb0.9     & 259,377,381                & 0.056                            & \textbf{175.6}                         & \textbf{20.3} $\pm$  0.7                    & \textbf{20.3} $\pm$  0.6                     & \textbf{22.4}  $\pm$  0.6                     \\
DiagT-rus-hb0.9 & 259,399,448                & 0.056                            & \textbf{40.1}                          & 17.0 $\pm$  1.0                   & 17.7 $\pm$  0.9                     & 19.8 $\pm$ 0.8                     \\
DiagT-rus       & 256,820,916                & 0.197                            & \textbf{13.86}                         & \textbf{22.6} $\pm$  2.3                    & \textbf{21.2}  $\pm$  1.05                   & \textbf{23.3} $\pm$  0.8                      \\
 
\bottomrule
\end{tabular}
\begin{tablenotes}
      \small
      \item Bold entries are DiagT-based results that are better than BR.
      \item Abbreviations: BR - binary relevance, DiagT - our proposed method, hb - hashing bucket ratio, rus - random under sampling, nnz - \# of nonzeros. Precision metrics are calculated using the three-fold evaluation with a 95\% confidence interval.
\end{tablenotes}
\end{table}

We show in Table ~\ref{table:results}  that DiagT and its variations obtain higher precisions in models DiagT, DiagT-hb0.9, and DiagT-rus and have faster execution time compared to BR. Model DiagT-rus-hb0.9, which employs both the hashing trick and undersampling suffers from a high bias. The hashing bucket ratio 0.9 is chosen after performing hyperparameter tuning.


\section{Conclusion}
We proposed a problem transformation method to solve multi-label learning via a single binary classification that is shown to have clear improvements in execution time and precision compared to the binary relevance method. In future work, we intend to perform a more extensive hyperparameter search, experiment with different dimensionality reduction techniques, and compare our method against other popular multi-label learning algorithms.

\bibliographystyle{plain}  

\bibliography{main}  

\begin{thebibliography}{1}

\bibitem{duda1973pattern}
Richard~O Duda and Peter~E Hart.
\newblock Pattern classification and scene analysis.
\newblock {\em A Wiley-Interscience Publication, New York: Wiley, 1973}, 1973.

\bibitem{har2003constraint}
Sariel Har-Peled, Dan Roth, and Dav Zimak.
\newblock Constraint classification for multiclass classification and ranking.
\newblock In {\em Advances in neural information processing systems}, pages
  809--816, 2003.

\bibitem{langford2007vowpal}
John Langford, Lihong Li, and Alex Strehl.
\newblock Vowpal wabbit online learning project, 2007.

\bibitem{nilsson1965learning}
Nils~J Nilsson.
\newblock {\em Learning machines: foundations of trainable pattern-classifying
  systems}.
\newblock McGraw-Hill, 1965.

\bibitem{tsoumakas2007multi}
Grigorios Tsoumakas and Ioannis Katakis.
\newblock Multi-label classification: An overview.
\newblock {\em International Journal of Data Warehousing and Mining (IJDWM)},
  3(3):1--13, 2007.

\bibitem{tsoumakas2009mining}
Grigorios Tsoumakas, Ioannis Katakis, and Ioannis Vlahavas.
\newblock Mining multi-label data.
\newblock In {\em Data mining and knowledge discovery handbook}, pages
  667--685. Springer, 2009.

\end{thebibliography}

\end{document}